\documentclass[10,2]{article}
\usepackage[a4paper, total={6.6in,9.8in}]{geometry}
\usepackage[utf8]{inputenc}
\usepackage{graphicx}
\usepackage{float}
\usepackage{subfig}
\usepackage{authblk}
\usepackage{xcolor}

\providecommand{\keywords}[1]
{
  \small        
  \textbf{\textit{Keywords---}} #1
}

\title{Ensuring the Robustness and Reliability of Data-Driven Knowledge Discovery Models in Production and Manufacturing}
\author[1,2]{Shailesh Tripathi \thanks{corresponding author} } 
\author[1]{David Muhr}
\author[1]{Brunner Manuel}
\author[2,3]{Frank Emmert-Streib}
\author[1]{Herbert Jodlbauer}
\author[1,4,5]{Matthias Dehmer}
\affil[1]{Production and Operations Management, University of Applied Sciences Upper Austria, Austria}
\affil[2]{Predictive Society and Data Analytics Lab, Faculty of Information Technolgy and Communication Sciences, Tampere University, FI-33101 Tampere, Finland}
\affil[3]{Institute of Biosciences and Medical Technology, FI-33101 Tampere, Finland}
\affil[4]{Department of Biomedical Computer Science and Mechatronics, UMIT-The Health and Life Science University, 6060 Hall in Tyrol, Austria}
\affil[5]{College of Artificial Intelligence, Nankai University, Tianjin 300350, China}
\affil[*]{corresponding author}
\date{}

\begin{document}

\maketitle
 \newpage
\onecolumn




\begin{abstract}
The implementation of robust, stable, and user-centered data analytics and machine learning models is confronted by numerous challenges in production and manufacturing. Therefore, a systematic approach is required to develop, evaluate, and deploy such models. The data-driven knowledge discovery framework provides an orderly partition of the data-mining
processes to ensure the practical implementation of data analytics and machine learning models. However, the practical application of  robust industry-specific data-driven knowledge discovery models faces multiple data-- and model-development--related issues. These issues should be carefully addressed by allowing a flexible,  customized, and industry-specific knowledge discovery framework; in our case, this takes the form of the cross-industry standard process for data mining (CRISP-DM). This framework is designed to ensure active cooperation between different phases to adequately address data- and model-related issues. In this paper, we review several extensions of CRISP-DM models and various data-robustness-- and model-robustness--related problems in machine learning, which currently lacks proper cooperation between data experts and business experts because of the limitations of data-driven knowledge discovery models.
\end{abstract}
\section{Introduction}
Since the beginning of industry 4.0 initiatives, the concept of smart manufacturing has gained considerable attention among researchers from academia and industry. Specifically, data-driven knowledge discovery models are now regarded as an essential pillar for smart manufacturing. The concept of intelligent manufacturing systems was initially discussed  in Refs. \cite{HATVANY_1978, HATVANY1983423}, where the authors discussed the prospects of systems and their complexities and emphasized that  systems should be built to be resilient to unforeseen situations \textcolor{black}{and to predict trends in real time for large amounts of data}.

In recent years, the idea \textcolor{black}{of smart manufacturing}  developed further with the framework of multi-agent systems (MASs). MASs are  groups of independent agents that cooperate with each other and are capable of perceiving, communicating, reproducing, and working not only toward a common goal but also  toward  individual objectives. An agent is composed of several modules that enable it to work effectively both individually and collectively. The acting module of a learning agent collects data and information (percepts) from the external world through sensors and responds through effectors, which results in actions. The learning module and critical module of an agent react to improve the actions and the performance standards by interacting with each other. Furthermore, the problem generator module enforces the exploratory efforts of the agent to develop a more appropriate world view \cite{MONOSTORI2003277, Lee_2008, KOUISS1997, WANG_2016, Russell_1995}. 

The learning agent is a software program or an algorithm that leads to an optimal solution to a problem.  The learning processes may be classified into three categories: (1) supervised,  (2) unsupervised, and (3) reinforcement learning. The learning module is a key driver of a learning agent and puts forward a comprehensive automated smart manufacturing process  for  autonomous functioning, collaboration, cooperation, and decision making. 
The availability and potential of  data from an integrated business environment allow  various business objectives to be formulated, such as automated internal functioning, organizational goals, and  social and economic benefits. Various business analytic methods, metrics, and machine learning (ML)\ strategies serve to analyze these business objectives. For instance,  Gunther et al. \cite{GUNTHER_2017191}  extensively reviewed  big data in terms of their importance to social and economic benefits. Furthermore, they highlight three main issues of big data in order to realize their potential and to match up with  the ground realities of  business functioning (i.e., how to transform data into information to achieve higher business goals). The three issues considered are work-practice goals, organizational goals, and supra-organizational goals. The availability of big data allows us to apply various complex ML models to various manufacturing problems that could not be addressed previously. However, big data do not provide  valuable insight into situations where the problem stems from not  developing sufficient data competencies or addressing organizational issues. Ross et al. \cite{Ross_2013} discussed the big data obsession of business organizations \textcolor{black}{rather than asking the straightforward question of whether big data are  useful and whether they  increase the value of  business-related goals}. They concluded that many business applications might not need big data to add value to business outcomes but instead need to address other organizational issues and business rules for evidence-based decision making founded on rather small data. However, their paper does not discuss  manufacturing and production but instead focuses on issues related to business management  rules. Nevertheless, a similar understanding may be obtained in many other production- and manufacturing-related instances that do not require big production data but instead require sufficient samples of observational or experimental data  to support robust data analytics and predictive models.     

 Kusiak \cite{Kusiak_2018}  discusses key features of smart manufacturing and considers predictive engineering as one of the essential pillars of smart manufacturing, and
Stanula et al. \cite{STANULA2018} discuss various guidelines to  follow for data selection as per the CRISP-DM framework for understanding business and data  in manufacturing.  Wuest et al.  \cite{Wuest2014}  discuss the various  challenges encountered in ML applications in manufacturing, such as  data preprocessing, sustainability, selection of the appropriate ML algorithm, interpretation of results,  evaluation, complexity, and high dimensionality. The challenges raised by Wuest et al. \cite{Wuest2014} require a systematic and robust implementation of each phase of the CRISP-DM framework, as per the manufacturing environment. Kusiak \cite{kusiak_2017} discusses the five key points that highlight the gaps in innovation and obstruct the advancement in smart manufacturing and  emphasizes that academic research must collaborate extensively to solve complex industrial problems.  The second point is about new and improved processes of data collection, processing,  regular summarization, and sharing. The third point is to develop predictive models to know outcomes in advance. The fourth point deals with  developing general predictive models to capture trends and patterns in the data to overcome future challenges instead of memorizing data by feeding  in  large amounts. The fifth point is to connect factories and processes that work together by using open interfaces and universal standards. Kusiak \cite{kusiak_2017} also emphasizes that  smart manufacturing systems should adopt new strategies to measure useful parameters in manufacturing processes to match up with new requirements and standards.
The studies discussed above emphasize the development of  ML models  and their robustness so that ML can effectively meet  the new manufacturing challenges. These robustness issues may be attributed to  faulty sensors, corrupt data, manipulated data, missing data, or data drifting, but not to the use of  appropriate ML algorithms. Here we highlight a set of common reasons that are responsible for the robustness of  ML models in manufacturing. 

The robustness of a model is   its effectiveness in accurately predicting  output when tested on a  new data set. In other words, the predictive power of a robust model does not decrease when testing it on a  new independent data set (which can be a noisy  data set). The underlying assumption used to build such a model is that the training and testing data have the same statistical characteristics. However, in a real-world production scenario, such models perform poorly on adversarial cases when the data is highly perturbed or corrupted \cite{kurakin2016adversarial}.
Goodfellow et al. \cite{Goodfellow_2018} discuss two techniques \textcolor{black}{for determining the robustness} of  ML models in adversarial cases: adversarial training and defensive distillation\textcolor{black}{. These techniques help  smooth the decision surface in adversarial directions.} Another approach to robust modeling is  model verification \cite{Goodfellow_2018, ranzato2019robustness}, which evaluates the upper bound of the testing error of ML models and thereby  ensures that the expected error does not exceed  the estimated error determined based on a wide range of adversarial cases. Model verification is a new field of ML research and certifies that a model does not perform worse in adversarial cases. \textcolor{black}{Various other ML\ approaches are still not widely applied; for example, some  studies  focus on deep learning models and support vector machines for image analysis.
}

In a production and manufacturing context, a robust ML model should not only perform well with training and testing data but should also  contribute to effective production planning and control strategies (PPCSs), demand forecasting, optimizing shop floor operations and quality improvement operations. Jodlbauer et al. \cite{Jodlbauer_2008} provided a stochastic analysis to assess the robustness and stability of PPCSs (i.e., the optimal parameterization of PPCS variables against environmental changes such as the variation of the demand pattern, machine breakdowns, and several other production-related issues). The robustness of PPCSs is the flexibility of adjusting PPCS parameters in a changing external environment without incurring additional cost, and the sensitivity of PPCS parameters are described as the stability.  Robust ML models can contribute to extracting deep insights from the data of several production processes and help to address various production and manufacturing goals, such as
\begin{itemize}
    \item [--] cost-effective production planning; 
    \item [--] improved production quality;
    \item [--] production in time;
    \item [--] long-term sustainability of production processes;
    \item[--] effective risk management;
    \item[--] real-time production monitoring;
    \item[--] energy efficient production \cite{GAHM_2016}, which   is another concern for green and environmentally friendly sustainable production. 
\end{itemize}

To develop robust ML models, we need to deal with various data and modeling issues, some of which are interrelated with each other. For this reason, we split these problems conceptually into two key points: (I) model selection and deployment-related issues and (II) data-related issues.
The model selection and deployment-related issues include
\begin{itemize}
\item model accuracy;
\item model multiplicity;
\item model interpretability;
\item model transparency.
\end{itemize} 
Data-related issues concern model robustness, which arises at different phases of the CRISP-DM framework, including
\begin{itemize}
 \item experimental design and sample size; 
 \item model complexity;
 \item class imbalance; 
 \item data dimensionality;
 \item data heteroscedasticity; 
 \item incomplete data in terms of variability; 
 \item unaccounted external effects;
 \item concept drift (when the data change over time); 
 \item data labeling;   
 \item feature engineering and selection.
 \end{itemize}
 
 
 In this paper, we address the model robustness issues of ML  models for data from manufacturing and production within the CRISP-DM framework. Specifically, data-driven approaches for knowledge discovery under the CRISP-DM framework need data analytics and ML models to apply, extract, analyze, and predict useful information from the data for knowledge discovery and higher-level learning for various tasks in production processes.

 Note that data acquisition for model building and deployment is not straightforward; it requires active cooperation between domain experts and data scientists  to exchange relevant information so that any need at the time of data preparation and data modeling can be addressed. Usually, the initial step is the business-understanding phase, which sets business goals, puts forward an execution plan, and  assesses the resources. Domain experts and higher-level management mostly set these objectives for the models, whereas  data scientists are involved in executing data-related goals and providing relevant data information in terms of scope, quality, and business planning.   {Different phases of CRISP-DM need active interactions between domain experts and data experts to obtain a realistic implementation of a data-driven knowledge discovery model. By ``active interactions,'' we mean  flexible cooperation between data experts and business experts, which allows the whole CRISP-DM process to work dynamically and updates the different phases of CRISP-DM. Such active interactions are allowed to develop new phases and interdependencies between different phases, thus ensuring a factory-specific data-driven knowledge discovery model.}
  The data-driven discovery models are most likely to encounter problems if  a lack of active cooperation exists between  groups. For example,
in the data-understanding phase, data experts (i.e., data engineers and data scientists) should  engage in active interactions, which helps to comprehensively understand the complexity of the problem  for executing data mining and ML projects. Data engineers should ensure data quality, data acquisition, and data preprocessing. However,  data scientists are needed to evaluate the model-building requirements and to select appropriate candidate models for achieving the project goals. Model evaluation and interpretation requires an equal contribution from domain experts and data scientists. Overall, this describes the CRISP-DM framework; see \textbf{Figure \ref{fig:crisp}}. The CRISP-DM framework provides a reasonable division of resources and responsibilities to implement knowledge discovery through data mining and ML.  The data-driven nature of CRISP-DM ultimately  maintains production quality, maintenance, and decision making.

Before discussing the fundamental issues of model robustness, we discuss in  more detail the CRISP-DM framework and model-robustness measures.

\section{Data Mining Process Models, Cross-Industry Standard Process for Data Mining}

\begin{figure}
    \centering
    \includegraphics[scale=0.65]{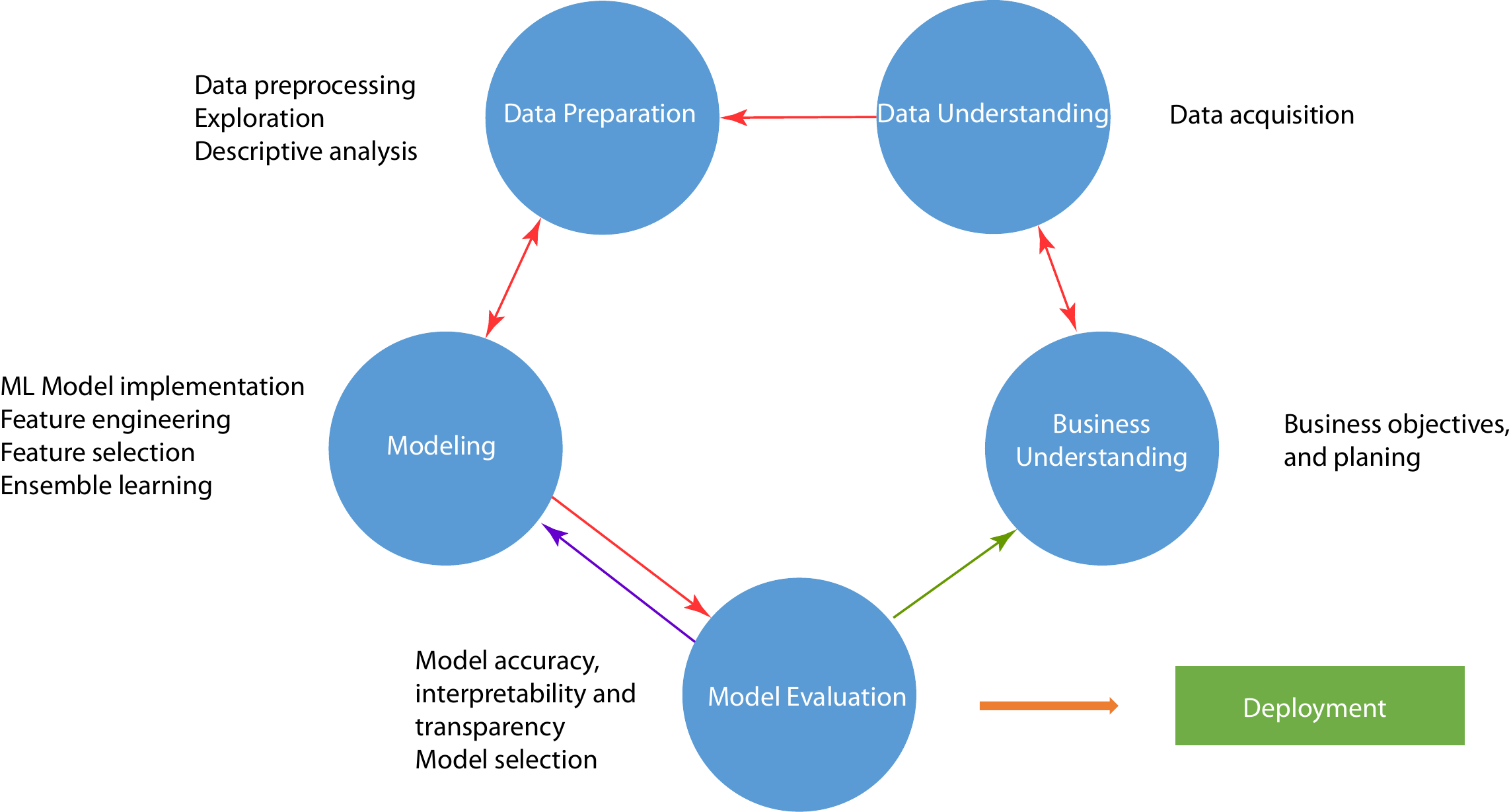}
    \caption{The CRISP-DM framework describes a cyclic process of a data analysis project. 
    }
    \label{fig:crisp}
\end{figure}

The CRISP-DM  framework was introduced in 1996. Its goal is to provide a systematic and general approach for applying data mining concepts to analyze business operations and gain in-depth insights into business processes \cite{crisp_2000}. It is a widely accepted framework for industrial data mining and data analytics for data-driven knowledge discovery. The CRISP-DM framework is an endeavor to provide a general framework \textcolor{black}{that is independent of any given industry and of applications that execute data mining methods that look at different stages of a business} \cite{crisp_2000}. Briefly, CRISP-DM divides data-mining-related knowledge discovery processes into six phases:
business understanding, data understanding, data preparation, modeling, evaluation, and deployment. \textbf{Figure \ref{fig:crisp}} shows a schematic overview of the CRISP-DM framework. 
{The arrows in the figure show the dependencies of different phases. For example, the modeling phase depends on data preparation and vice versa. When the data preparation phase is complete, the next step is the modeling phase, for which different candidate models with different assumptions about the data are proposed.  To meet the data requirements for model building,  data experts might step back to the data preparation phase and apply various data-preprocessing methods.
Similarly, in another example, the data-acquisition and data-preprocessing steps are done by data experts so that   business goals can be supplied with the available data. In such cases, data quality, data size, and expected features based on prior business understanding should be present in the data. Business experts and data experts can agree to obtain the data through a specific experimental design or by acquiring new observational data.  

Furthermore,  business goals need to be readjusted 
based on an understanding acquired during the data preparation and data understanding phases. The readjustment of business goals may imply the required time, cost, data quality, and the usefulness of the business plan. Once the data experts and business experts  agree on data issues, the data analysis process moves forward to model development and evaluation; similarly, these phases require cooperation between different groups.}
Overall, CRISP-DM describes a cyclic process, which has  recently been highlighted as the emergent feature of data science \cite{em_jJuly010115}. In contrast, ML or statistics focus traditionally on a single method for the analysis of data. Thus, the CRISP-DM framework follows a data science approach.

A recent example of such an approach is the Data mining methodology for engineering applications (DMME) \cite{HUBER_2019}
model, which is an extension of the CRISP-DM model for engineering applications. This model adds three new phases, namely, technical understanding, technical realization (which sits between business understanding and data understanding), and technical implementation (which sits between the evaluation and deployment phases). The DMME model 
also draws new interactions between different phases, namely, \textcolor{black}{between evaluation and technical understanding, between data understanding and technical understanding, and between technical understanding and technical realization}. Furthermore, this model also emphasizes the cooperation between specialized goals for the refinement of the framework. 

In  light of big data, technical advancements, broader business objectives, and advanced data science models, Grady \cite{grady_2016} discussed the need for a new, improved framework.

He coined the term ``knowledge discovery in data science'' (KDDS) and discussed different aspects of KDDS, such as KDDS phases, KDDS process models, and KDDS activities. Grady also emphasized that this approach establishes a new, improved framework, which he calls the cross-industry standard process model for data science (CRISP-DS). Table \ref{tab:crisp_extension} lists the extensions for the CRISP-DM model.  These models show that the basic CRISP-DM framework  cannot  satisfy the variety of data mining requirements from different sectors. 

\begin{figure}
    \centering
    \includegraphics[scale=0.45]{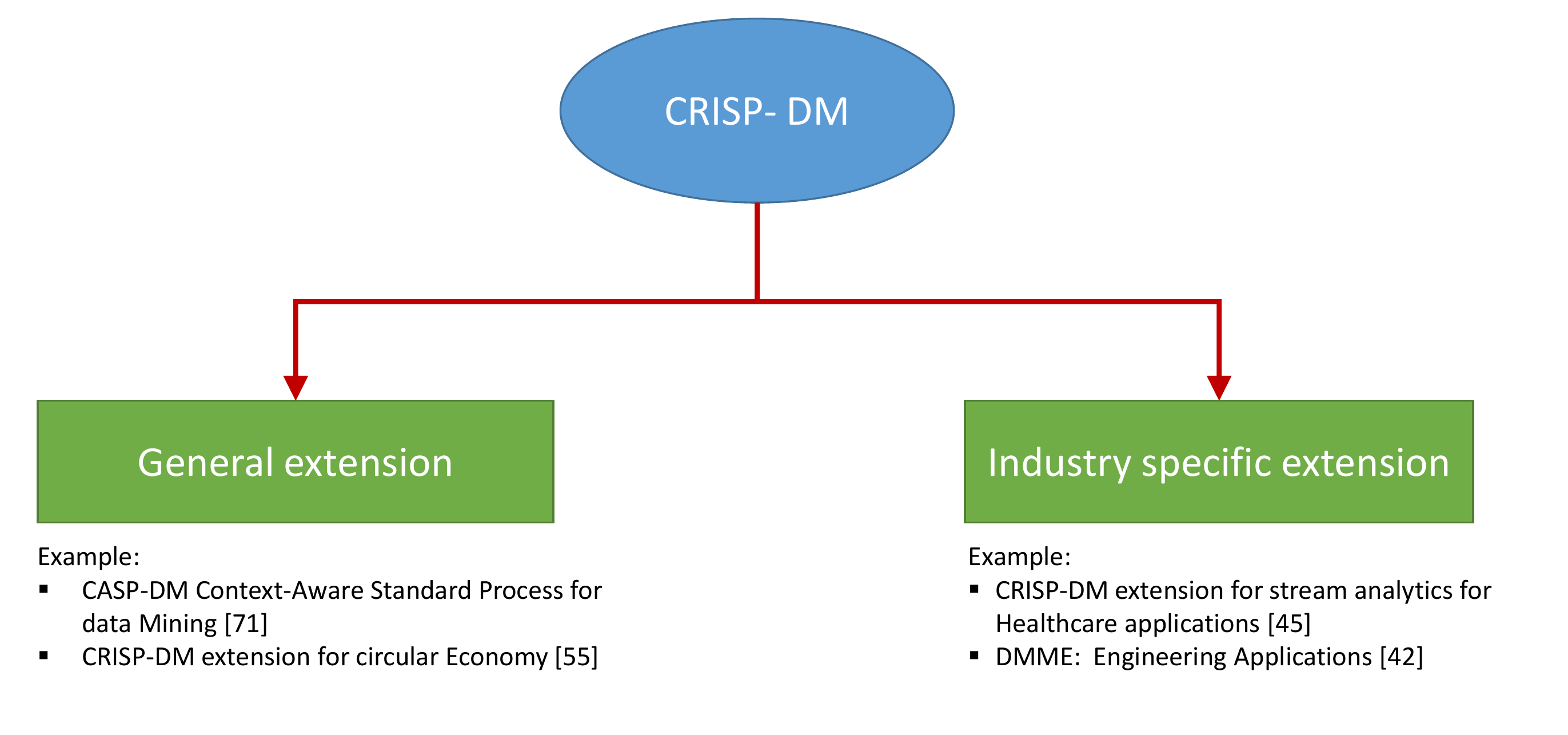}
    \caption{Classification of CRISP-DM framework into two classes: general extension and industry-specific extension. }
    \label{fig:crispextension}
\end{figure}

We now highlight two types of CRISP-DM data mining models: the industry- or application-specific extension, and the concept-based extension (see \textbf{Figure \ref{fig:crispextension}}). For the circular economy, KDD and CRISP-DM  are the concept-based evolution of the CRISP-DM model.  However, DMME and CRISP-DM for stream analytics are application-specific customizations of the basic model. Given the new industrial trends, concept-based evolution has become a generalized extension of CRISP-DM. For example, the circular economy is a growing trend in the industry for lifelong services, product upgrades, and product recycling \cite{Stahel_2016} for sustainable and environmentally conscious manufacturing.  
All business sectors should gradually adopt the circular economy concept. Therefore, a knowledge discovery model must adopt concept-related extensions of the CRISP-DM framework.

To implement a robust and flexible DM framework in a company, one needs to elaborate on the different phases of CRISP-DM and its extended structures because new technologies and emerging trends in production and manufacturing demand advanced and flexible data mining and knowledge discovery models.

\begin{table}[!h]
    \centering
\caption{List of extensions of  CRISP-DM process model framework based on  industry-specific requirements (application-specific requirements) due to changing business trends.}
    \label{tab:crisp_extension}
    \begin{tabular}{|p{4.0cm}|p{6.5cm}|p{4cm}|}
    \hline
    Model & Description & Application \\
    \hline
     DMME \cite{HUBER_2019}  & Adds two new phases between business understanding and 
     data understanding and one phase between model evaluation and deployment& Engineering applications\\
     \hline
     KDDS with big data \cite{grady_2016} & Adding various new activities, especially to handle big data  
     &  A proposed framework as a need for the current scenario in big data and data science applications\\
     \hline
     CRISP-DM extension for stream analytics \cite{Kalgotra_2016} & Data preparation and data modeling stage to be redefined for 
     multidimensional and time-variant data, where the IoT system sends multiple signals over time  & Healthcare application.\\
     \hline
     CRISP-DM extension in context of circular economy \cite{Kristoffersen_2019} &  Adds a data validation phase and new interactions between different phases& Aligning business analytics with the circular economy goals  \\
     \hline
     Context-aware standard process for data mining (CASP-DM) \cite{Martnez_2017} &The deployment  context of the model can  differ from the training context. Therefore, for  context-aware ML models and model evaluation, new activities and tasks are added at different phases of CRISP-DM.  & Robust and systematic reuse of data transformation and ML models if the context changes \\
      \hline
      APREP-DM \cite{Nagashima_2019} & Extended framework for handling outliers,  missing data, and data preprocessing at the data-understanding phase & General extension for automated preprocessing of sensor data\\
      \hline
      QM-CRISP-DM \cite{Schafer_2018} & CRISP-DM extension for the purpose of quality management considering DMIAC cycle & Adding quality management tools in each phase of CRISP-DM framework validated in the real-world electronic production process \\  
      \hline
      ASUM-DM \cite{haffar_2015} & IBM-SPSS 
      initiative for the practical implementation of the CRISP-DM framework, which combines traditional and agile principals. It implements existing CRSIP-DM phases and adds additional operational, deployment, and project-management phases. & General framework that allows comprehensive, scalable, and product-specific implementation.\\
      \hline
      A variability-aware design approach for CRISP-DM \cite{Vale_2018}& Extends the structural framework to capture the variability of data modeling and analysis phase as feature models for more flexible implementation of data process models& General framework which considers model and data variability for the improved automation of data analysis\\  
      \hline
    \end{tabular}
\end{table}

\section{Human-Centered Data Science}

The new wave of artificial intelligence (AI) and ML solutions aims to replace numerous human-related tasks and decision-making processes by automated methods. This raises concerns regarding responsible and ethical ML and AI models that do not underestimate human interests and do not incorporate  social and cultural biases. The intended objective of AI is to help humans  enhance their understanding of larger system-related goals (e.g., societal goals).  The ML systems have, until now,   primarily focused on applying  cost-effective automations that are useful to business organizations for maximizing profits.
Such automations are not part of complex decision-making processes. However, in the future, autonomous systems would take over for many crucial decision-making processes. 
One  example of such an automation  is self-driving cars.
Many large car manufacturing companies have announced the launch of self-driving cars, but the acceptability and use of such vehicles is a major challenge \cite{Howard2014PublicPO, Urmson_2008}. 
Kaur and Rampersad \cite{KAUR_201887}   quantitatively analyze  the acceptance of technology that addresses human concerns. The study analyzes the public response to the issues, which are security, reliability, performance expectancy, trust or safety, and adoption scenarios.

\begin{figure}[!t]
    \centering
    \includegraphics[scale=0.38]{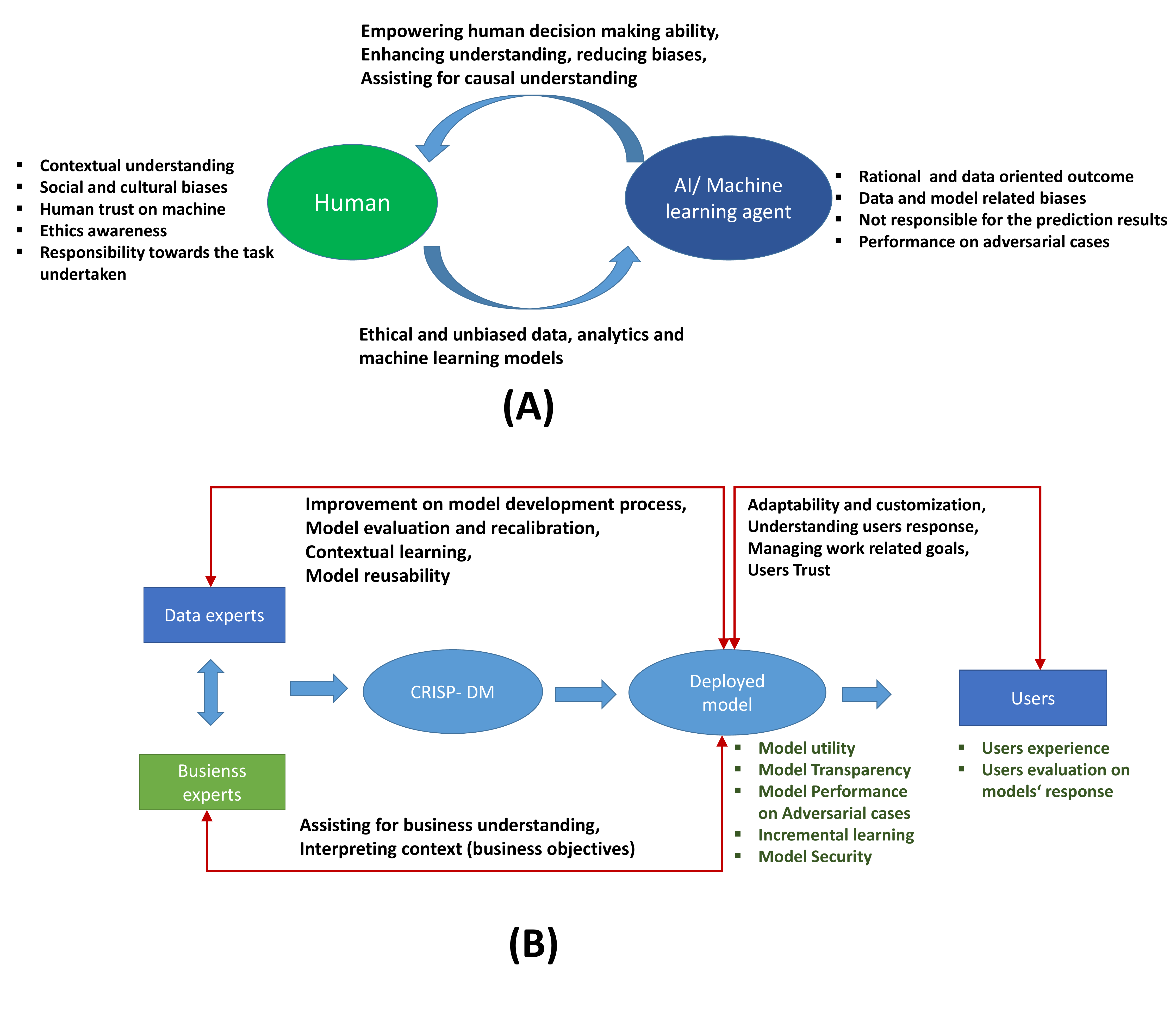}
    \caption{\textbf{(A)} Schematic view of interactions between humans and AI and ML agents. \textbf{(B)} Schematic view of post-deployment of  ML model not left in isolation but continuously updating and addressing human concerns by allowing human interactions to ensure a human-centered data science.}
    \label{fig:postcrisp}
\end{figure}

Similarly, such hypotheses can be discussed in industrial production and manufacturing cases \textcolor{black}{when a self-automated system or a self-functioning machine based on AI or ML systems decides various production and manufacturing tasks}.

In these cases, how a system would function in complex scenarios would be a question, both ethical and technical.

This type of autonomy can affect the complex production process of PPCSs estimation, which might, in adversarial cases,  influence the robustness and stability of  PPCSs, predictive maintenance, and other production processes \cite{Fradinho_2019}. Another futuristic question one can ask is, what should be done with an automated system when it is deployed? Should it be left with no human intervention, should it always  supersede human understanding, and how can humans and AI systems cooperate effectively? 

In industrial production  cases, human-centered data science issues can be seen from the following human-centric perspective:
\begin{itemize}

    \item [--] what is the role of human-centered ML and data science processes in decision making related to work specific goals, business goals, and societal goals?
       
    \item [--] can production processes be completely automated with no involvement of humans by AI agents with human-like intelligence? 
    
    \item [--] the emergence of complexity when a series of tasks is automated and integrated: can an AI or ML system  exhibit the higher level of intelligence required for independent and integrated decision making?

    \item [--] when industrial production processes are completed by a combination of humans and ML agents, how can humans and ML\ agents  interact effectively for integrated analysis and decision making?

\end{itemize}

The first point mentioned here is still in its infancy and awaits the future of integrating  complex data mining and ML processes. The future design of data science modeling of complex data and business goals is beyond the scope of this paper. However, a fundamental question that remains  relevant is how can large-scale automation led by AI and ML models impact  humans? At its core, this question probes how AI can efficiently serve,  empower, and facilitate human understanding and decision-making tasks  and provide  deeper insight into the solutions to these complex problems. This question also leads to the second and third points; namely, should we leave the decision making entirely in the hands of AI-ML models? In other words, can a machine decide by itself  what is right or wrong for humans?  Complete automation would lead to the emergence of many complex aspects of the production process, and dealing with such complexity would be the future challenge of research into production and manufacturing. 
 In such complex cases, the role of humans  should be to interact and collaborate with the automated systems. They should use intelligent models  for decision making and have the wisdom to override machine decisions in exceptional cases. 

The human perspective and machine and statistical learning have different characteristics and capabilities. A ML model makes decisions based on a narrow data-related view with no  contextual understanding. The AI-ML models cannot be held responsible for their decisions. However, they can produce results rationally and logically based on the data used to train them. Interactions between humans and machines are possible in two ways: The first way involves the model developer and ML models, where the training of models are the responsibility of data experts and business experts. They should train models with a diverse range of data with no   bias and no  breach of data ethics. 
The second way is between model users and ML models. The results predicted by models empower and assist humans to make decisions  regarding  simple to complex tasks.   \textbf{Figure \ref{fig:postcrisp}(A)} shows a schematic view of human-machine interactions. In this figure, we highlight various characteristics of human and ML models and how effective interactions between humans and ML models complement and empower human decision making and further improves the capacity of models to make fair   analyses.

 Further extending the human-machine interaction concept, we now discuss the industrial-production framework shown in \textbf{Figure \ref{fig:postcrisp}(B)}, which highlights a CRISP-DM process in the deployment phase of the model.
The deployed model should have proven its utility and  allow transparency so that the user can understand the models' prediction and trust  the results. 
The deployed model interacts with data experts, business experts, and actual users. In an industrial production process, the users are technical operators or machine handlers.  These interactions have different meanings for different users. Data experts interact with the model to improve the  prediction accuracy and model performance. They provide contextual meaning to the results predicted by the model and train it further for contextual learning. The data experts also explore the reusability of a model for other cases \cite{pan_2010}.
Business experts should test these results against their understanding of  larger business goals and should  explain the business context based on model results. Furthermore, the results should allow the model to be reused and adaptable to changes in production and manufacturing scenarios.  

In production and manufacturing, data-driven knowledge processes must adopt human-centered concerns in data analytics and ML models; the human-machine interaction must not fizzle out after the deployment of the model \cite{amershi2014power}. 
The model should remain  interactive with its users and be allowed to  evolve and update itself. The other characteristics of the deployed models are their transparency and performance in adversarial cases. Model transparency should enable users to access various model features and evaluate the predicted results.
Additionally, the human-machine interaction must be aware of the surrounding environment so that the model can adapt to  environmental changes, as required for  intelligent learning of ML models. Continuous interactions between humans and machines improve the generality of the model by incremental learning and update the model so that it gives a robust and stable response and  can adapt to changes.

\section{Model Assessment}

Model building and  evaluation in CRISP-DM are two phases  in which data are (i) searched for patterns and (ii)  analyzed. These models are divided into four classes: descriptive, diagnostic, predictive, and prescriptive  \cite{Bertsimas_2020}. The primary goal of these models is pattern recognition, machine-health management, condition-based maintenance, predictive maintenance, production scheduling, life-cycle optimization, and supply-chain management  prediction.  Alberto et al. \cite{Alberto_Diez_2019} provide a comprehensive review of various methods applied in manufacturing and production with representative studies of  descriptive,  predictive, and prescriptive models of various industrial sectors, including manufacturing. Vogl et al. \cite{Vogl2019}  review  current practices for diagnostic and prognostic analysis in manufacturing and highlight the challenges faced by the prognostic health management  system in the current scenario of automated maintenance.   

\begin{figure}
    \centering
    \includegraphics[scale=0.45]{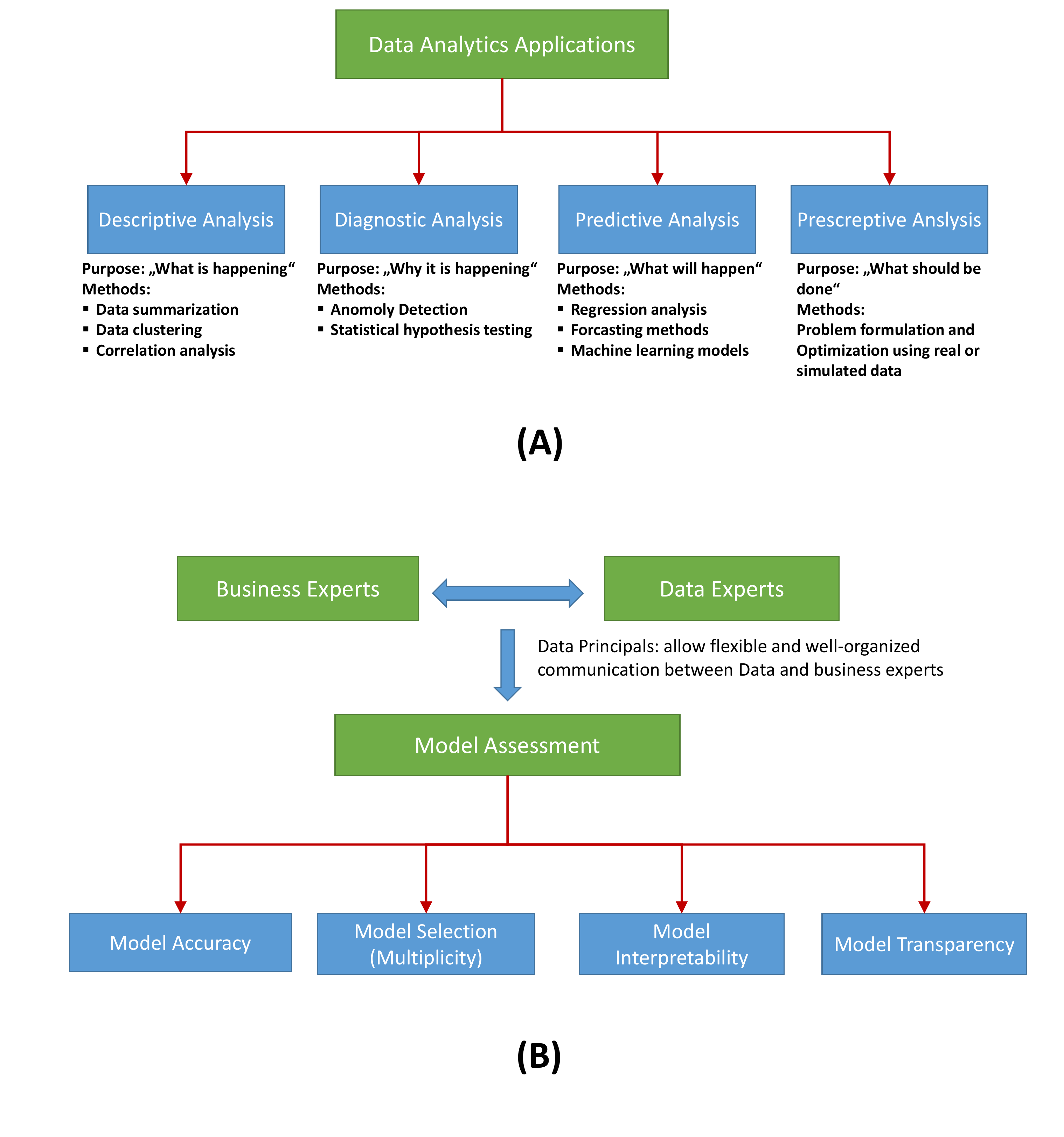}
    \caption{\textbf{(A)} Components of data analytics problems. \textbf{(B)} Model assessment.}
    \label{fig:app}
\end{figure}

\textbf{Figure \ref{fig:app}(A)} shows a schematic view of the division of data analysis problems. The divisions are useful to understand the nature of the problem and the methods that are applicable to the problem.

\begin{figure}
    \centering
    \includegraphics[scale=.65]{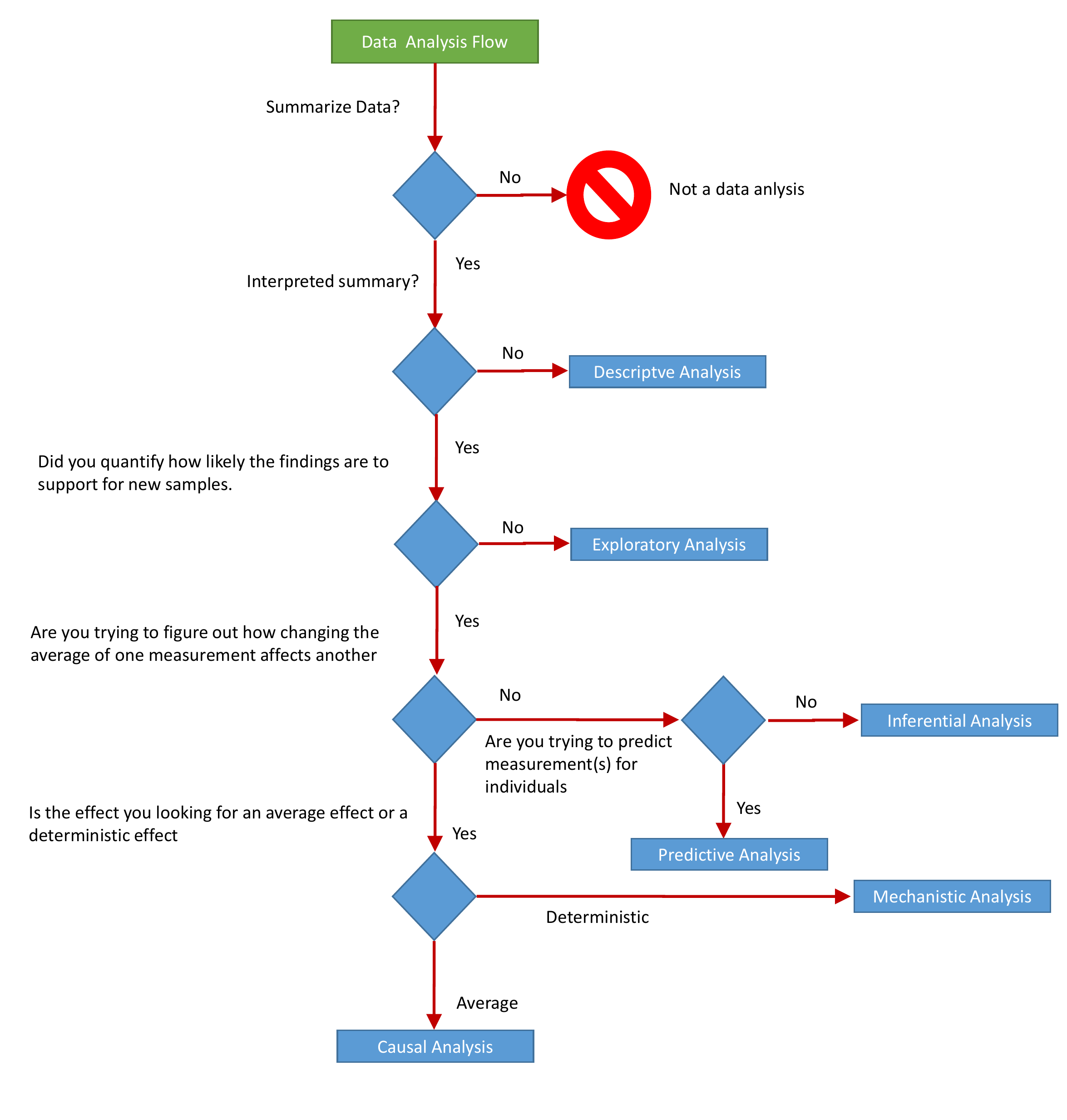}
    \caption{Schematic diagram showing  data analytics process. Connections are shown between the individual analysis steps that constitute the whole project \cite{Leek2015}.}
    \label{fig:flowchart}
\end{figure}

Leek \cite{Leek2015} emphasizes that, even if the statistical results are correct,  data analysis can be wrong because  the wrong questions are asked. One must be aware of such a situation in an industrial framework and should select the right approach for the analysis. He further divides data analysis into six categories (see \textbf{Figure \ref{fig:flowchart}}). This chart is useful for the basic understanding of data analysis for business experts and data experts, which helps prevent  knowledge discovery from diverging from the real objective. 

The business understanding and the later  data modeling approaches for  solving the type of problem shown in \textbf{Figure} \textbf{\ref{fig:app}A} requires a systematic interaction between data experts and business experts. Such communication is possible only when the data experts and business experts are aware \textcolor{black}{of each other} and agree with each other. Therefore, they need to address the data- and model-related communication in the context of business understanding.
Hicks et al. \cite{hicks_2019} discuss the vocabulary required   to convey the accurate meaning of the data analysis. Additionally, They describe the six principles of  data analysis: Data matching, Exhaustive, Skeptical, Second order, Transparent, and Reproducible. 
These  six principles are applicable to an industrial framework to communicate the results of the data analysis. A clear understanding of these principles helps business experts and data experts  develop robust models for data-driven knowledge discovery. 
Data- and model-related issues are not  restricted only to understanding  business  and data; the most crucial part is model assessment, which requires accurate nonambiguous model selection, interpretability, and transparency. Model assessment depends on various issues, namely, model accuracy, model multiplicity, model interpretability, and model transparency (see schematic diagram  in \textbf{Figure \ref{fig:app}B}),  and various strategies can be applied to obtain the best model assessment.   Model assessment requires an agreement between data and business experts, so all four components of model assessment shown in \textbf{Figure \ref{fig:app}B} should be addressed properly. 
 The first part is the model accuracy; Raschka \cite{Sebastian_2018}  reviews  model evaluation, selection, and ML model selection for various ML methods, and   Palacio-Ni\~ no  and Berzal \cite{palacionio2019}   review  the evaluation of unsupervised learning methods. 
Akaike information criterion (AIC), Bayesian information criterion (BIC), cross-validation error, and probabilistic model selection are some of the approaches used to select  less complex models.
  Model interpretation is another crucial issue in data-driven knowledge discovery with ML models.

In general,  monotonically increasing functions (linear and nonlinear), such as additive models, sparse linear models, and decision trees, are regarded as interpretable models, whereas  nonlinear and nonmonotonic functions have higher complexity and thus  a lower interpretability \cite{hall_2018, ribeiro2016modelagnostic}.
Doshi-Velez  \cite{doshivelez2017rigorous} provides a theoretical framework for the interpretability of models by   defining three classes for evaluating  interpretability:
The first class is application-grounded evaluation, where the domain experts evaluate the model with real tasks. Next, human-grounded evaluation allows a model to be tested with simplified tasks without a specific end-goal. The third class is  functionally grounded evaluation, which uses a predefined definition of interpretability of a model as proxy and optimizes the candidate models. In an industrial setup, most of the model evaluations involve functionally grounded evaluation. Application- and human-grounded evaluation requires more resources and could be  expensive and time consuming. \textcolor{black}{Functionally grounded evaluation can be useful for  learning  if the selection of proxy models is based on factors that are relevant to real-world applications}. 

Model agnostic interpretability is another approach for interpreting ML models. In this approach, a model is treated as a black box, and the model is analyzed by entering perturbed input data and analyzing the  respective output \cite{ribeiro2016modelagnostic}. The model agnostic approach allows the application of a wide range of black-box ML models as interpretable for predictive analysis. Local interpretable model-agnostic explanations  \cite{ribeiro2016i} explain a model locally by constructing a local binary classifier of a regularized linear model  by using simulated or  perturbed input data to explain an instance predicted by the black-box machine learning model.

Another issue with  model assessment is the multiplicity of  models; such cases are known as the ``Rashomon'' effect \cite{breiman_2001}, whereby multiple predictive models make predictions with competing accuracy of the same target variable. In such cases, one should not come to a conclusion about the validity of a prediction in terms of its explanatory variables from one or a few models until the hypothesis of model multiplicity is  invalidated. In industrial scenarios, the Rashomon effect can occur when we have multiple models with competing accuracy;  one can then choose a narrative to interpret the results based on the selected model. Such interpretations can  differ for different models and would impact the business objective or business understanding. Fisher \cite{fisher_2018models} proposes the model class reliance, which is a measure of a set of variables of a model that show high predictive accuracy. Marx  et al. \cite{marx2019predictive} propose the  ambiguity and discrepancy measures to evaluate the multiplicity problem in classification models.  In this approach, models are selected from a set of ``good'' models, which maximizes the discrepancy compared to a proposed baseline model. To  select a model with simplicity and accuracy, Semenova
and Rudin \cite{semenova_2019study} propose the {\it{Rashomon ratio}} measure, which is a ratio of competing models  in  hypothesis space.  


In a business scenario, when manufacturing becomes increasingly complex and integrated,  data-driven models are useful to make informed decisions and automate functioning, which allows for  efficient and cost-effective production.  Even for a model with high accuracy, user trust is always a crucial consideration for the deployment of a model. In such cases, transparency is always  a matter of concern and should be addressed with simple explanations.
Trust issues arise because of  data-source uncertainty, the performance of deployed models, and user belief in the model output. This uncertainty can seep into the whole process of data mining, from data preprocessing to model building, and thus impacts the entire business-intelligence process.
Sacha et al. \cite{Sacha_2015} provide a framework to address human-machine trust issues that are caused by uncertainties propagated in data processing,  model building, and visual analytics. The proposed framework to deal with the uncertainty is divided into two parts: The first part is the system-related uncertainty and emphasizes  handling machine uncertainty by quantifying uncertainty in data processing and model building, aggregation of total uncertainty at different phases of data analysis, and interactive exploration of uncertainty through visual and other means. The second part involves  human factors that emphasize transparency, which allows  system functions to access  experts and users for review and to explore uncertainty, thus building an awareness of the emerging uncertainty---and additionally tracking human behavior, which is also useful to estimate  human bias and how it impacts  human-machine interactions.

Data transparency and model transparency should ensure that the entire process is explainable to business experts and  users. It should also be secure from external attacks, which allows models to be deployed for making business decisions, for user control, and for social acceptance of intelligent systems \cite{Aaron2019, weller2017transparency}. Weller \cite{weller2017transparency} discusses eight types of transparency rules for developers, business experts, and users to help them  understand the existing system functioning. Transparency allows a large and complex system to be robust and resilient and self-correcting, and ultimately ensures steady and flexible models.


\section{Data-related robustness issues of machine-learning models}

A combination of data understanding, business understanding, and data preparation are the steps that contribute to model building. A comprehensive CRISP-DM setup should  address various business-related issues and ultimately contribute to business understanding. Data modeling and evaluation are central units in a CRISP-DM framework. {In the following, we discuss data issues for robust data analytics and ML models. The shortcomings of data, such as sample size, data dimensionality,  data processing, data redundancy, feature engineering, and several other data problems, should be addressed with the appropriate strategy for  robust ML models. Several data-related issues are interdependent with data understanding, \textcolor{black}{such as the preparation and modeling phases of the CRISP-DM framework}. In such cases, any data-related issues reflect on model evaluation and on the deployment phase in terms of underperformance and biased predictions of real-world problems.} 
  
\subsection{Experimental design and sample size}
In  data-driven process optimization, the data may be observational data or  experimental data. Experimental data are used to test the underlying hypothesis to build an understanding and to optimize the response variable by controlling or modifying various combinations of input parameters. In  manufacturing, the experimental designs  mostly focus on optimizing  various parameters  for quality control. \textcolor{black}{The experimental design is a key criterion for determining whether  a model employs all the right answers to be understood in ML-based analysis and manufacturing optimization}. The production process is controlled by setting various process variables, which, during production,  are  monitored for  changes. In an experimental design, we monitor certain process variables by controlling other variables to understand how they affect product quality. The one-factor approach is  common and involves repeating  experiments  with the factors changed between each experiment until the effect of all factors is recorded. However, this one-factor approach is time consuming and expensive. For a robust experimental design, Taguchi proposed a methodology that uses  orthogonal arrays, signal-to-noise ratios, and the appropriate combination of factors  to study the entire parameter space through a limited number of experiments \cite{taguchi_1987, taguchi_1989, Unal_1991, Chang_2001}. An optimal experimental design is a useful strategy that stands between business and data understanding. This approach also benefits model robustness in the evaluation phase in the CRISP-DM framework. The active interaction  between business understanding and data understanding serves to solve the optimal sample-size issue and prevent an underpowered analysis.

\subsection{Model complexity}
The complexity of predictive models is sensitive to the prediction of testing data.  Any complex model can lead to  overfitting, whereas simpler models lead to prediction bias. This property of predictive models is known as ``bias-variance trade-off.'' Complexity in the model could lead to  robustness issues, such as large testing error or the wrong interpretation of model parameters. In some cases, if  future data contain a lower signal-to-noise ratio,  large prediction error may occur.  This problem can arise because of a large number of correlated features or through feature engineering, where we create a large number of redundant features with high correlation or when we try to fit a high-degree polynomial model in our data set.

Vapnik–Chervonenkis (VC)
dimension \cite{vapnik_1971} is a quantitative measure for measuring the complexity of  models and is used to estimate the testing error in terms of model complexity. A model is selected based on the best performance with testing data and, as the complexity increases, the optimized training error decreases and the expected testing error first increases, then decreases, and then increases again. The training error and VC are used to calculate the upper bound of the testing error. This method is called  ``structural risk minimization'' and is a model that minimizes the upper bounds for the risk. 
Regularization \cite{Chang_2001,Tibshirani_1970,yuan_2006, frank_2019} and feature-selection methods (both wrapper- and filter-based) \cite{KOHAVI1997273,MALDONADO20092208} are the useful strategies of variable selection to keep  model complexity in check by minimizing bias and variance. Hoffmann et al. \cite{HOFFMANN201550} propose a sparse partial robust \(M\) regression method, which is a useful approach for regression models to remove  variables that are  important due to the outlier behavior in the data. 
The model complexity affects the modeling and model evaluation, such as model description, robustness, and parameter settings in the CRISP-DM framework.

\subsection{Class imbalance}
In  production and manufacturing data,  class imbalance is a common phenomenon. One such example is  vibration data from  cold testing of engines: all manufactured engines go through cold testing, and the technical expert tries to identify different errors in the engines by using strict threshold criteria. They label any errors found with a particular error label based on their technical knowledge. The proportion of  engines with errors and the number of good engines are very low. To implement a multiclass classification model using  previous data, which automatically identifies the error class of an engine, developing such models with high accuracy is difficult. Skewed sample distributions are a common problem in industrial production-related quality control, and imbalanced data lead to  models where predictions are inclined toward the major class. In such cases, the major class is ``good engines.'' If a model classifies a faulty engine  as a good engine (high false positives), then it would severely affect the reputation of quality control. In such cases, false positives could severely affect  product reputation. Undersampling of major classes and oversampling of minor classes form the basis of the solution commonly proposed for the problem. References \cite{Tajik_2015, DUAN2016239,ZHANG201556} discuss    fault detection models with an imbalanced training dataset in the industrial and manufacturing sector. Resampling methods (oversampling, undersampling, and hybrid sampling) \cite{Chawla_2002, han_2005, NEKOOEIMEHR_2016, SUN_2015, CATENI_2014}, feature selection and extraction, cost-sensitive learning, and ensemble methods \cite{HAIXIANG_2017, Krawczyk_2016} are other approaches to deal with the class-imbalance problem. The use of evaluation measures considering the presence of undersampled classes is also recommended to evaluate  models, such as the probabilistic thresholding model \cite{su_2007}, adjusted \(F\) measure \cite{MARATEA_2014}, Matthews correlation coefficient \cite{MATTHEWS_1975}, and AUC/ROC.
The imbalance is critical to  model building and  evaluation problems in the CRISP-DM framework.

\subsection{Data dimensionality} 
In ML and data mining,  high-dimensional data contain a large number of features (dimensions), so the requirement of optimal samples increases exponentially. Therefore, a high-dimensional dataset  is sparse \((p\gg n)\). 
In  modern  manufacturing, the complexity of the manufacturing processes is continuously increasing, which also increases the dimensionality of process variables that need to be monitored. Wuest et al. \cite{Wuest2014} discuss  manufacturing efficiency and product quality by considering the complexities and high dimensionality of manufacturing programs.
High-dimensional data are a challenge for  model robustness, outlier detection \cite{zimek_2012}, time, and algorithmic complexity in ML. These higher-dimensional data require changes in existing methods or the development of new methods for such sparse data. One approach is to use a dimensionality-reduction strategy such as \textcolor{black}{a principal component analysis } \cite{SUBASI20108} and use the principal components for ML methods. Such a strategy also requires a careful examination because of  information loss when data are projected onto lower dimensions. Similarly, another strategy is feature selection, where  the most important feature variables are found and we remove the redundant features from the model. Regularization \cite{Tibshirani_1970},  tree-based models \cite{HSU_2004, NIPS2017_6907,Bertsimas2017}, mutual information \cite{Vergara_2014, BENNASAR_2015},  and clustering are other approaches for feature selection.

\subsection{Data heteroscedasticity}

In regression problems, the underlying assumption \(\hat y\) for the error term in the response data  is that it has constant variance (homoscedasticity). Although such assumptions are useful to simplify models,  real-world data do not follow the assumption of the constant variance error term and thus violate the assumption of homoscedasticity. Therefore, real data are  heteroscedastic. Predictions based on a simple linear model of such data would still be consistent but can lead to practical implications that produce  unreliable estimates of the standard error of coefficient parameters,
thus leading to bias in test statistics and confidence intervals.  Heteroscedasticity can be caused  directly or indirectly by effects such as  changes in the scale of observed data, structural shifts in the data and outliers, or the omission of explanatory variables.   Heteroscedasticity can be parametric, which means that it can be described as a function of explanatory variables \cite{white_1980},  or unspecified, meaning that it cannot be described by explanatory variables.  Examples in manufacturing are related to predictive maintenance (i.e.,  predicting the remaining useful life of a machine based on vibration data), quality control, and optimization  \cite{lee_2019, tamminen_2013}.  Heteroscedasticity in regression models is identified by analyzing residual errors. Residual plots and several statistical tests may be used to test for heteroscedasticity in  data \cite{GODFREY_1978, white_1980, Breusch_1979, Carol_1990}.  Other studies and methods  to model the variance of response include weighted least squares \cite{white_1980}, the heteroscedastic kernel \cite{CAWLEY_2004}, heteroscedastic Gaussian process regression \cite{Kersting_2007, KOU2015298}, heteroscedastic Bayesian additive regression trees (BART)
\cite{Matthew_2017}, and heteroscedastic support vector machine regression \cite{Hu_2016}. 

\subsection{Incomplete and missing data}

Data quality is an essential part of  data-driven process optimization, process quality improvement, and smart decision making in production  planning and control. 
In the manufacturing process, data come from various sources and is heterogeneous in terms of variety and dimensionality. Data quality is one of the important issues for the implementation of robust ML methods in process optimization.  Data quality may suffer for various reasons such as missing values, outliers, incomplete data, inconsistent data, noisy data, and unaccounted data \cite{KOKSAL_2011}. The typical way to deal with  missing data is to delete them or replace them with the average value or most frequent value and apply multiple imputations \cite{fritz_2005} and expectation maximization \cite{allen2010}. Another method is to consider the probability distribution of missing values in statistical modeling \cite{Rubin_1976,little_2019}. These considerations are of three types: (1) missing completely at random, (2) missing at random, and (3)  missing not at random.  Joint modeling and fully conditional specifications are two approaches to multivariate data imputation \cite{Buuren_2006,Buuren_2007}.  Recent studies to impute missing data in production and manufacturing and optimize  production quality considered classification and regression trees  to impute manufacturing data \cite{white_2018}, modified  fully conditional specifications \cite{tong_2017}, and multiple prediction models of a missing data set to estimate product quality  \cite{seokho_2018}. Loukopoulos et al. \cite{Loukopoulos_2018}  studied  the application of different imputation methods  to industrial centrifugal compressor data and compared several methods. Their results suggest multiple imputations with self-organizing maps and \(k\) nearest neighbors as the best approach to follow. Incomplete and missing data have repercussions in business and data understanding, exploratory analysis, variable selection, model building, and evaluation. A careful strategy involving imputation or deletion should be adopted for missing data. 

\subsection{External effects}

External effects are important factors in data-driven process optimization. In a big data framework,  external effects might be unnoticed, incorrectly recorded, or not studied in detail (i.e., unknown). External effects can involve environmental factors (e.g., temperature, humidity), human resources, machine health, running time, material quality variation, and other factors that are dealt with subjectively but  significantly affect production and data quality. For example, to create a ML model to predict product quality, the data (process parameters) are  recorded by sensors from several machines. In the injection molding process, the process parameters cannot estimate product quality alone because product quality is significantly affected by temperature, humidity,  machine health, and material type, and these external factors must be considered to optimize quality.
Similarly, in the sintering process \cite{strasser_2019}, where product quality is predicted in terms of its shape accuracy, the  data from various experiments show a considerable variation in different process parameters from unknown and unaccounted sources. In such cases, the prediction can be biased or  erroneous with high variance. A careful examination involving exploratory analysis is required to detect such effects. Gao et al. \cite{GAO_2017} studied a hierarchical analytic process to produce a comprehensive quality evaluation system of large piston compressors that considers external and intrinsic effects.

\subsection{Concept drift}
Changes in data  over time due to various factors such as external effects, environmental changes, technological advances, and other various reasons emerge  from unknown real-world phenomena.  ML models, in general, are trained with static data and thus do not consider the dynamic nature of the data, such as changes over time in the underlying distribution. In such case, the performance of these models deteriorates over time. Robust ML modeling requires  identifying changes in data over time, separating noise from  concept drift,  adapting changes, and updating the model.   
Webb et al. \cite{Webb2016} provide a  hierarchical classification of concept drift that includes five main categories: drift subject, drift frequency, drift transition, drift recurrence, and drift magnitude.
They make the case that, rather than defining  concept drift qualitatively, a quantitative description of concept drift is required to understand problems in order to detect and address time-dependent distributions. These quantitative measures are  cycle duration, drift rate, magnitude, frequency, duration, and path length and are required for a detailed understanding of concept drift.  Such an understanding  allows  modeling by new ML methods that are robust against various types of drifts because they can update themselves by active online learning and thereby detect drifts.
Lu et al. \cite{lu_2018} provide a detailed review of current research, methods, and techniques to deal with problems related to concept drift. One  key point in that paper is that most of the methods identify ``when'' a drift  happens in the data but cannot answer ``where'' or ``how'' it happened. They further emphasize that research into concept drift  should consider the following themes:
\begin{itemize}
\item [(1)] identifying drift severity and regions to better adapt to concept drift; 
\item[(2)] techniques for unsupervised and semisupervised drift detection and adaptation;
\item[(3)] a framework for selecting real-world data to evaluate  ML methods that handle concept drift;
\item[(4)] integrating ML methods with concept-drift methodologies
(Gama et al. \cite{Gama2013} discuss the various evaluation and assessment practices for concept-drift methods and strategies to  rapidly and efficiently detect concept drift).
\end{itemize}
Cheng et al. \cite{cheng_2019} present the ensemble learning algorithm for condition-based monitoring with concept drift and imbalance data for offline classifiers. Yang et al. \cite{yang_2019} discuss a novel online sequential extreme learning model  to detect various types of drifts in  data. 
Finally, Wang and Abraham \cite{Heng_2015} proposed a concept-drift detection framework known as ``linear four rates'' that is applicable for batch and streaming data and  also deals effectively with  imbalanced datasets.  

In the manufacturing environment,  concept  drift is an expected phenomenon. When discussing with  technical experts, data drift should be understood and explained clearly by experts for data and business understanding to ensure  that future models are flexible and  adaptable to changes due to data drift.

\subsection{Data labeling}

Data annotations (i.e., labels)  as output allow ML methods to learn  functions from the data. Specifically,  input data are modeled through  class labels by using supervised ML models for classification problems.
If the labeling is noisy or not appropriately assigned, the  predictive performance of the classification model is severely degraded. 
Therefore,  for supervised ML models, data labeling or annotation is a nontrivial issue for data quality, model building, and evaluation. The labeling can  either be done manually by crowd sourcing or by using semisupervised learning, transfer learning, active learning, or probabilistic program induction \cite{ZHOU_2017}. Sheng et al.  \cite{Sheng_2008} discuss repeated-labeling techniques to label noisy data. In a typical manufacturing environment,  labeling is done by operators or technical experts; in some cases, the same type of data is annotated by multiple operators who work in different shifts or at different sites. Sometimes, the operators do not follow a strict protocol and  label intuitively based on their experience. In such cases, data quality  can suffer immensely. Inadequate learning from data may occur if   a large number of samples is not available for a specific problem and the data labeling is noisy.  

{Data labeling  is a part of the data understanding phase and should follow a clear and well-defined framework based on the technical and statistical understanding of the manufacturing and production processes and their outcome. Many manufacturing and production tasks are managed by machine operators, technical experts, engineers, and domain experts using measures and experience to optimize the quality of the output. Because these experts are well aware of the existing situation, they should be allowed   to label data to produce robust ML models and maintain data quality.}

\subsection{Feature engineering}

The deep learning framework provides a promising approach for automated smart manufacturing. These algorithms undertake automatic feature extraction and selection at higher levels in the networks. They are used for numerous important tasks in manufacturing, such as fault detection \cite{LU_2017} or predicting machine health  \cite{david_2017}. Wang et al. \cite{WANG_2018}  comprehensively reviewed  deep learning methods for smart manufacturing. Importantly, deep learning models may suffer from the same issues as discussed in previous sections. Data complexity, model and time complexity, model interpretation, and the requirement of large amounts of data means that they  may not be suited for every industrial production problem. Therefore, the demand remains strong in industrial production for traditional ML algorithms using feature-selection methods. 

In ML problems, a feature is an attribute used in supervised or unsupervised ML models as an explanatory or input variable. Features are constructed by applying mathematical functions to raw attributes (i.e., input data) based on domain knowledge. As a result, feature engineering and extraction lead to new characteristics of an original data set (i.e., it constructs a new feature space from the original data descriptors). Features can be created by methods that are linear, nonlinear, or use independent component analysis.
The main objective of feature-extraction techniques is to remove redundant data and  create interpretable models to improve prediction accuracy or to create generalized features that are, e.g., classifier independent. In general, feature learning enhances the efficiency of regression and classification models used for predictive maintenance and product quality classification. In the manufacturing industry, feature engineering and extraction is a common practice for the classification of engine quality, machine health, or similar problems using vibration data. For instance, fast Fourier transform  
or power spectral density  statistical features are standard feature extraction methods used for vibration data \cite{Caesarendra_2017}.  

Sondhi \cite{Sondhi_2009} discusses various feature-extraction methods  based on decision trees, genetic programming, inductive logic programming, and annotation-based methods. Interestingly, genetic programming in combination with symbolic regression \textcolor{black}{searches for a mathematical equation that best  approximates a target variable}. Given its simple and understandable structure, this approach has been useful for many industrial applications of modeling and feature extraction    \cite{strasser_2019, YANG_2015, Kotanchek_2010} . However, training for a large number of variables, model overfitting, and the time complexity of the model are problematic for this approach.
Caesarendra and Tjahjowidodo \cite{Caesarendra_2017} discuss five feature categories  for vibration data: 
\begin{itemize}
\item[1.] phase-space dissimilarity measurement;
\item[2.] complexity measurement;
\item[3.] time-frequency representation;
\item[4.] time-domain feature extraction;
\item[5.]  frequency-domain feature extraction. 
  \end{itemize}
{These feature engineering methods are useful to summarize various characteristics of vibration data and reduce the size of high-dimensional data. Mechanical vibration is a useful indicator of machine functioning. The features of vibration data are often used to train ML models to predict fault detection, do predictive maintenance, and for various other diagnostic analyses.} 
He et al. \cite{HE_2019} discuss the statistical pattern analysis  framework for fault detection and diagnosis for batch and process data. In this approach, instead of process variables, sample-wise and variable-wise statistical features that quantify process characteristics are extracted and used for process monitoring. This approach is useful for real-world dynamic, nonlinear, and non-Gaussian process data, which are transformed into multivariate Gaussian-distributed features related to process characteristics that are then used for  multivariate statistical analysis.   
Ko et al. \cite{Ko_2017} discuss various feature-extraction and -engineering techniques for anomaly detection or binary classification in different industries.

Recent results for deep learning networks and support vector machines  demonstrate that feature selection can  negatively affect  the prediction performance for high-dimensional genomic data \cite{em_jsmolander2019comparing}. However, whether these results translate  to data in other domains remains to be seen. Until then, feature selection remains a standard method of data preprocessing.

In general, for the CRISP-DM framework, feature engineering is essential for real-world problems of model building. Real-world data features are not independent---they are correlated with each other, which hints of  interactions between the underlying processes in manufacturing. Many of the higher-level effects related to the maintenance and product quality are  combinations of lower-level effects of several processes and are reflected by the data. Therefore, approaches that use feature engineering, feature extraction, and feature selection  are useful to build models to predict or test for higher-level effects.

\section{Conclusion}

This paper addresses some of the key issues affecting the performance, implementation, and  use of data mining and ML models. Specifically, we   discuss  existing methods and future directions of model building and evaluation in the area of smart manufacturing. In this context, CRISP-DM is a general framework that provides a structured approach for data mining and ML applications. 
Its  goal is to provide comprehensive data analytics and ML toolboxes to address the data analysis challenges in manufacturing and ultimately to help  establish standards for data-driven process models that can be implemented for improving production quality, predictive maintenance, error detection, and the other problems in industrial production and manufacturing. 

The CRISP-DM framework describes a cyclical flow of the entire data mining process with several feedback loops.
However, in many realistic data mining and modeling cases, it is an incomplete framework that requires additional information and active interactions between different sections of CRISP-DM.
The demand for additional or new information can arise due to real-world production complexity  (e.g., data-related or industry-specific applications, other organizational and manufacturing issues, or new emerging business trends).

This review discusses data and model-related issues for building robust models that are successively linked to a business understanding. One of the challenges for  robust model building is the active interaction between different phases of the CRISP-DM framework. For instance, business understanding and data understanding go hand in hand because we learn business objectives (business understanding) and use relevant data (data understanding), so data should be explored appropriately (data preparation) by various statistical and analytical tools to enhance our business understanding from a realistic viewpoint. These three phases require  active cooperation by data experts and business experts so that any gap or misinformation can be quickly corrected.
Similarly, each phase in the CRISP-DM framework can seek interactions through a feedback loop to  clarify  the process, and such interactions can be temporary or  permanent extensions of the model. Thus, the active interactions between the different phases of the CRISP-DM framework should be systematically and efficiently managed so as to improve the whole process and make it  easy to customize and update. 

Overall, CRISP-DM describes a cyclic process that is similar to general data-science approaches \cite{em_jJuly010115}. Interestingly, this  differs from ML or statistics that focus traditionally on a single method only for the analysis of data, establishing in this way one-step processes. Thus, CRISP-DM can be seen as an industrial realization of data science.

Another important issue  discussed herein is  related to the model assessment from the viewpoint of data experts, business experts, and users. We discuss four essential components of model assessment---model accuracy, model interpretability, model multiplicity, and model transparency---which are crucial for the final deployment of the model. A model should not deviate from its  goal as it encounters new data; it should predict results with accuracy and provide a satisfactory interpretation. Multiple models with competing accuracy  also present a challenge when it comes to selecting a robust and interpretable model.

We  define cases where the interpretability of the model is crucial for business understanding and decision making.  The fourth issue is model transparency, which should allow  users to obtain information on the internal structure of algorithms and methods to enable them to use various functional parameters so that the deployment phase can be implemented and integrated with the least effort and  users can use the model and rely on its predictions.   

This review focuses on the following important points for the application of data-driven knowledge discovery process models:
\begin{itemize}
    \item [a.] We adopt a relevant and practically implementable, scalable, cost-effective, and low-risk CRISP-DM process model with industry-specific and general industrial-trend related extensions. 
    \item[b.] Each phase of the CRISP-DM framework should interact actively and self-actively so that any data and model issue can be resolved with no lag.
    \item[c.] When using CRISP-DM, we define the problem category related to industrial production and maintenance for the practical implementation of data analysis.
    \item[d.] Define the type of analysis to be performed for the model-building to avoid wrong answers to the right questions.
    \item[e.] Various data issues should be appropriately handled for data understanding and business understanding, such as exploration, preprocessing, and model building, which are discussed in Section \(4\).
    \item[f.] Model evaluation and deployment should ensure that  the implemented model satisfies the requirements of model accuracy, model interpretability, model multiplicity,  and model transparency.
    {\item [g.] The CRISP-DM framework should adopt a human-centric approach to allow for transparent use of the model and so that users and developers can evaluate, customize, and reuse the results that enhance their understanding of  better production quality, maintenance, and other production, business, and societal goals.}

\end{itemize}

In general, for data-driven models, data are at  center stage and all  goals are realized by the potential of the available data \cite{em_jemmert2019defining}. The assessment of achieved business goals at the end phase of the CRISP-DM is also analyzed through the success of the model. Thus, all  newer understanding  depends on the data  obtained and how we turn that data into   useful information.  These  objectives cannot be obtained solely through the efforts of business experts or data experts. Instead, a collective effort of active cooperation is required for the appropriate process  of human-centered CRISP-DM to achieve larger business goals. 
{Additionally, a human-centered approach should be considered for the integrated and complex system where we deploy several ML models to monitor and execute multiple production and manufacturing tasks. These various models and the human business experts, data experts,  technical experts, and users cannot be independently functioning entities. They interact with each other and form a complex system that  leads to the emergence of various higher-level characteristics of smart manufacturing. In such complex systems, the various AI and machine tasks must be transparent, interactive, and flexible so as to create a robust and stable manufacturing and production environment.}
Future  research in this direction will explore  industry-specific requirements to develop   data-driven knowledge discovery models that can be implemented in practice.

\bibliographystyle{plain}
\bibliographystyle{plain}
{\scriptsize{
\bibliography{ref,bib_ref_journals}
}}

\end{document}